\input{aipcheck}

\documentclass[
    ,final            
  ]
  {aipproc}

\layoutstyle{6x9}

\def\gta{\ifmmode {\mathbin{\lower 3pt\hbox   
    {$\,\rlap{\raise 5pt\hbox{$\char'076$}}\mathchar"7218\,$}}}
    \else {${\mathbin{\lower 3pt\hbox
    {$\rlap{\raise 5pt\hbox{$\char'076$}}\mathchar"7218\,$}}}
    $}\fi}
\def\lta{\ifmmode {\,\mathbin{\lower 3pt\hbox   
    {$\,\rlap{\raise 5pt\hbox{$\char'074$}}\mathchar"7218\,$}}}
    \else {${\mathbin{\lower 3pt\hbox
    {$\rlap{\raise 5pt\hbox{$\char'074$}}\mathchar"7218\,$}}}
    $}\fi}

\begin{document}

\title{Formation Scenarios for Intermediate-Mass Black Holes}

\author{M. Coleman Miller}{
  address={University of Maryland, Department of Astronomy,
College Park, MD  20742--2421}
}

\begin{abstract}
Black holes with hundreds to thousands of solar masses are
more massive than can be formed from a single star in the
current universe, yet the best candidates for these objects
are not located in gas-rich environments where gradual
accretion could build up the mass.  Three main formation scenarios
have been suggested in the literature: that intermediate-mass
black holes are the remnants of the first, metal-poor, stars;
that they result from direct collisions in young stellar clusters;
or that they are produced by gradual interactions and mergers
of compact objects in old dense clusters.  We discuss
each of these in turn and speculate on future observations that
may help sharpen our understanding of the formation of
intermediate-mass black holes.
\end{abstract}

\maketitle


\section{Introduction}

Black holes are solidly established to exist in the mass ranges $\sim
5-20\,M_\odot$ in our Galaxy and others (stellar-mass black holes) and
$\sim 10^{6-10}\,M_\odot$ in the centers of many galaxies (supermassive
black holes). As discussed by Mushotzky and by van der Marel in these
proceedings, there is now growing evidence for the existence of black holes
in the mass range $10^{2-4}\,M_\odot$ (intermediate-mass black holes, or
IMBH), especially in dense stellar clusters.  Such objects would be strong
sources of gravitational waves in a unique frequency range.  The rates of
merger events, as well as the information that could be gleaned from IMBH
in binaries, depend on the mechanism by which the IMBHs were formed.

When considering how IMBH may be formed, one can adapt Shakespeare,
as suggested by Keith Arnaud: ``Some are born great, some achieve
greatness, and some have greatness thrust upon 'em.''  Specifically,
it may be that the objects we now see as IMBH were born with approximately
their current mass, or it may be that through accretion or mergers,
black holes with initially much smaller masses grow to their current
size.  We now consider each of these options briefly, then go into them
in more detail in the following sections.

First, suppose that a black hole of mass $\sim 10^{2-4}\,M_\odot$ was
born with that mass.  We assume that the initial creation of a black hole
is always through the collapse of the core of a massive star.  Therefore,
if a single star leaves behind an intermediate-mass black hole, the star
itself obviously had to have at least as much mass as the remnant black
hole.  As stars with masses $M\gta 200\,M_\odot$ are thought not to form
in the current universe (for qualitative arguments, see Larson \&
Starrfield 1971), black hole remnants with this mass are ruled out.
Instead, the very early universe, where metallicities were probably small
enough that cooling was minimal and pulsational instabilities were weak,
might have produced stars with the requisite mass (e.g., Abel et al.
1998; Bromm, Coppi, \& Larson 1999; Bromm et al. 2001; Abel, Bryan, \&
Norman 2000; Fryer, Woosley, \& Heger 2001; Schneider et al. 2002;
Nakamura \& Umemura 2002).  These so-called Population III stars are
therefore candidate progenitors for intermediate-mass black holes.

If instead IMBH were grown from a smaller seed, we can narrow down the
ways in which it acquired mass.  In general, acquisition of mass can
take place by accretion or by mergers.  Consider accretion.  A black
hole in the interstellar medium will gain mass by Bondi-Hoyle accretion
of the medium.  However, the rate of accretion is tiny, leading to
growth timescales of (see, e.g., Miller \& Hamilton 2002a)
\begin{equation}
M/{\dot M}_{\rm B-H}\approx 10^{13}(M/10\,M_\odot)^{-1}(\rho/10^{-24}~
{\rm g~cm}^{-3})^{-1}(v/10^6~{\rm cm~s}^{-1})^3~{\rm yr}\; .
\end{equation}
Here we assume that the interstellar medium has density $\rho$ and
thermal velocity $v$ at infinity relative to the black hole.  The
shortest timescales would exist for cool, dense, molecular clouds, but 
even then the
accreting matter is pre-heated by the accretion luminosity (e.g.,
Maloney, Hollenbach, \& Tielens 1996; compare Blaes, Warren, \& Madau
1995 for accretion onto neutron stars), and the timescale is still
billions of years, much longer than either the lifetime of a molecular
cloud or the time for a black hole to cross the cloud (see Miller \&
Hamilton 2002a). Therefore, accretion from the interstellar medium is
insufficient unless gas is funneled to the hole, as may happen in the 
centers of galaxies via bar instabilities but not at the off-center
locations of IMBH.

The only way to accrete mass quickly enough is via accretion from stars
or mergers with stars or compact objects.  However, since individual
stars or stellar-mass compact objects are themselves much less massive
than the eventual IMBH, many such mergers or accretion events are
necessary.  In the disk of a galaxy, encounters with stars are far too
rare to account for the required change in mass.  Only in a dense stellar
cluster can there be multiple encounters as needed.  The cluster could be
a young stellar cluster, where interactions with massive main
sequence stars dominate, or an old cluster such as a globular cluster,
where interactions with compact remnants are most important.

In summary, the three ways currently considered to make IMBH are
as remnants of individual Population III stars, as the result of stellar
interactions in a young cluster, or as the result of interactions with
compact objects in an old cluster.  We now consider these possibilities
in turn.

\section{Population III Stars}

Stars that form in the current universe have masses limited by two
effects.  First, even if the Jeans mass of a molecular cloud is large,
metal line cooling is efficient enough that as portions of the cloud
contract they fragment into regions of mass $M\lta 100\,M_\odot$.
Second, although one might imagine that additional accretion could
push the mass arbitrarily high, radiation forces and pulsational
instabilities at  $M\gta 100\,M_\odot$ are thought to exist that would
expel matter faster than it could accrete (e.g., Larson \& Starrfield
1971).  In addition, even stars that do form with $M\sim 100\,M_\odot$
are thought to leave behind black holes of much smaller masses,
because of mass loss due to stellar winds (Fryer \& Kalogera 2001).

All of these issues might be circumvented if the metallicity is
sufficiently low, as it would be for stars formed in an environment of
primordial composition.  Cooling is then limited by rotational
transitions of molecular hydrogen.  It is therefore possible that
many of the first stars formed at hundreds or even thousands of solar
masses (e.g., Abel et al. 1998; Bromm, Coppi, \& Larson 1999; Bromm et
al. 2001; Abel, Bryan, \& Norman 2000; Schneider et al. 2002; Nakamura \&
Umemura 2002).  Moreover, given that both pulsational instabilities and
wind losses are driven by radiation forces on metal lines, these may be
insignificant for metal-free stars (Fryer, Woosley, \& Heger 2001).

Even so, not all Population III stars will leave behind black holes of
hundreds of solar masses.  The fate of the first generation of stars has
been explored by Nakamura \& Umemura (2001) and others, and although
there is still substantial uncertainty it is thought that the remnant
depends on mass in a relatively straightforward way.  If the initial mass
is $10\,M_\odot\lta M_{\rm init}\lta 40\,M_\odot$, there is likely to be
a standard core-collapse supernova that leaves behind a black hole with a
mass $\sim 5-20\,M_\odot$. If $40\,M_\odot\lta M_{\rm init}\lta
100\,M_\odot$, it is believed that the energy transferred to the stellar
envelope by the core collapse is insufficient to unbind the envelope,
hence the mass of the resulting black hole is close to the mass of the
original star.  If $100\,M_\odot\lta M_{\rm init}\lta 250\,M_\odot$,
however, another process enters.  As discussed by many authors (e.g.,
Barkat, Rakavy, \& Sack 1967; Bond, Arnett, \&
Carr 1984; Glatzel, El Eid, \& Fricke 1985;
Heger \& Woosley 2002), at these masses the overburden of
matter in the core is such that, in order to provide enough pressure,
oxygen burns at a temperature $kT\gta m_ec^2/3$.  At such temperatures
there is pair production.  The pairs are nonrelativistic and therefore
provide little pressure, hence the core has to contract further, raising
the temperature and increasing the number of pairs.  This process runs
away and causes the fusion of $\sim 40\,M_\odot$ of oxygen within a short
time, completely disrupting the star and leaving no remnant.  Only when
$M_{\rm init}\gta 250\,M_\odot$ is the binding energy of the envelope
enough to withstand the pair production instability, so a star of this
mass again leaves behind a black hole with a mass close to that of the
original star.  Therefore, the abundance of IMBH from Population III
stars depends entirely on the fraction of the first stars with initial
masses more than $\approx 250\,M_\odot$.

This fraction is still uncertain.  Cooling is likely to be dominated by
rotational transitions of molecular hydrogen.  However, since H$_2$ is
homopolar, its lowest-order rotational transition is quadrupolar, and
occurs at a comparatively high temperature of $T=510$~K.   If HD is present
in sufficient quantities, then its nonzero dipole moment and larger mass
allows transitions at $T=128$~K, which could lower the fragmentation mass
by a factor of several (Nakamura \& Umemura 2002).  On the other hand, if
there are extra sources of ionization present (e.g., an active galactic
nucleus) then H$_2$ may be dissociated, leading to much higher temperatures
and therefore, potentially, stars in the thousands of solar masses (e.g.,
Barkana \& Loeb 2001; Mackey, Bromm, \& Hernquist 2003). It is therefore
not clear at this time how many stars with $M_{\rm init} \gta 250\,M_\odot$
form in a given galaxy, although progress in this field has been rapid and
consensus may be achieved in the next few years. It is also not clear where
such objects would tend to form.  In standard hierarchical assembly models,
large dark matter halos are formed by the merger of many smaller halos, so
one might expect that black holes from Population III stars (likely formed
in high density halos) would congregate in the centers of large galaxies,
and possibly merge.  In this sense, it may be somewhat surprising that many
ultraluminous X-ray sources are found in globular clusters around the
elliptical galaxy NGC~1399 (Angelini, Loewenstein, \& Mushotzky 2001).
However, there are abundant unknowns about the mass and spatial
distribution of Population III remnants, so this is a viable model for
IMBHs.

\section{Interactions in Dense Clusters}

As discussed in the introduction, significant growth of a black hole
requires residence in a dense stellar cluster if the black hole is not
at the center of a galaxy.  In addition, as shown by Mushotzky and
by van der Marel in these proceedings, the best candidates for IMBH are
observed to be in clusters currently, regardless of their origin.  The
dynamics of clusters are therefore essential to the understanding of
intermediate-mass black holes, and especially to their potential as
sources of gravitational radiation.  Here we will first discuss general
dynamical effects, then examine separately young and old clusters.

\subsection{Dynamics of stellar clusters}

Stellar clusters of the mass and density of globular clusters or young
super star clusters (mass $\sim 10^{5-6}\,M_\odot$, number density
$\sim 10^{5-6}$~pc$^{-3}$ in the center) are wonderful testbeds for
dynamics.  This is because, unlike the central bulges of galaxies,
stellar clusters have evolution timescales significantly less
than their ages.  The relevant timescale is the relaxation time 
(e.g., Binney \& Tremaine 1987, pg. 190)
\begin{equation}
t_{\rm rel}\approx \left(N\over{8\ln N}\right)t_{\rm cross}\; ,
\end{equation}
where there are $N$ stars in the cluster and the crossing time is
$t_{\rm cross}$.  For example, the relaxation time in the core of
a globular is $t_{\rm rel}\sim 10^{7-9}$~yr, compared to its
$\sim 10^{10}$~yr age.

In a cluster with components of a single mass, the tendency is for the
core to contract while the outside of the cluster expands.  In this as in
many other ways, there are productive analogies with thermodynamics. For
example, the contraction and expansion are driven by an increase in the
total entropy of the cluster because the extra phase space available to
the outer stars more than compensates for the decreased phase space in the
core.  When the contrast between central density and density at the edge
reaches a critical value $\sim 700$, then single pointlike Newtonian objects
undergo a rapid ``core collapse" in which the central density formally
becomes infinite in finite time (see Binney \& Tremaine 1987, \S8.2, for
a discussion).  Clearly, other physics will
intervene, and the accepted primary physics is the interactions of
primordial binaries, which we discuss below.

In real multimass clusters, thermodynamic equilibrium would require that
the temperature of all components be equal, which for a cluster implies
that the average kinetic energy ${1\over 2}m_i\langle v_i\rangle^2$ is
the same for all components $i$.  In a gravitational field, smaller speed
means that a star will sink, hence more massive objects congregate
towards the center. In some situations, the more massive objects can
decouple from the lighter stars, in a process called the Spitzer
instability (Spitzer 1969).  In any case, the centers of dense clusters
are expected to be enriched in massive stars and binaries (e.g., Spitzer
\& Mathieu 1980), because on average binaries have more mass than single
stars.

The interactions of binaries have a profound impact on the dynamics of
dense clusters.  The effective cross section of interaction of a binary
is close to the area of the binary orbit, hence binaries are much
more interactive than single stars. Over the course of
many interactions, sufficiently wide binaries  (called ``soft'') are
widened more and more by interactions with single stars until they
eventually are separated into single stars (a process called
ionization).  Soft binaries have little net effect on the cluster.
However, sufficiently tight binaries (called ``hard'') tend to be
tightened by interactions with single stars.  The tendencies for soft
binaries to soften and hard binaries to harden are called ``Heggie's
laws'' (Heggie 1975) and have been explored numerically in many investigations.
For many mass ranges (but not all; see Quinlan 1996), an approximate
condition is that when in the three-body center
of mass frame the total energy (kinetic plus potential) is positive, the
binary is soft, but if the total energy is negative, the binary is hard.
Some of the tendencies in binary-single interactions are depicted in
Figure~1.

\begin{figure}
  \includegraphics[height=.3\textheight]{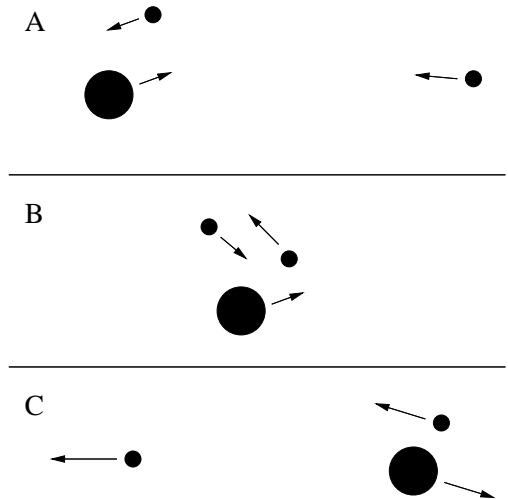}
  \caption{Typical interaction between binary and single star. In
Panel~A, the binary (at left) is approached by a single star from the
right, which is initially unbound to the binary.  We assume here that
the system has negative total energy as measured in the barycentric
frame.   In a close encounter, as in Panel~B, the interactions can be
extremely complicated and last for hundreds of orbits or more.  However,
in a Newtonian point mass  interaction without the possibility of
dissipation, the system must eventually resolve itself into a binary and
a single star, as in  Panel~C.  For a hard binary, the likely result (as
shown here) is that the binary has tightened, and that the binary
consists of the two most massive of the three original star. The
distribution of the fractional change in semimajor axis is independent
of the original semimajor axis for very hard binaries, hence the recoil
kick becomes larger for tighter binaries.}
\end{figure}

Hard binaries are extremely important in clusters.  The hardening in an
interaction produces recoil, hence binary-single interactions act as a
heat source in the centers of clusters, which stabilizes the core against
collapse for as long as the binaries can be tapped for energy (see Rasio,
Fregeau, \& Joshi 2001 for a recent investigation).  Even if all stars are
initially single, when the density becomes high enough ($\sim
10^{7-9}$~pc$^{-3}$ for central velocity dispersions $\sim
10-20$~km~s$^{-1}$; Goodman \& Hut 1993, Lee 1995) three initially
mutually unbound stars can interact in such a way as to bind two of them
together.  In addition, non-pointlike stars (especially giants) can
dissipate energy tidally, creating binaries during close passes of other
stars.  However, it is believed that the most important binaries are primordial,
formed at the origin of the cluster (Goodman \& Hut 1989). These provide
significant heating when the binaries are hardened (because hardening is
accompanied by recoil), at densities close to the observed
$10^{5-6}$~pc$^{-3}$.  This explains why some 20-40\% of globulars are
formally poised at the edge of core collapse (see Pryor \& Meylan 1993 for
data on globulars), despite this phase being very short-lived if all the
stars are single.  Over the long term, all the primordial binaries in a
cluster will tighten enough that they no longer interact significantly and
the cluster will collapse to much higher densities $\sim 10^8$~pc$^{-3}$,
but this is likely to take much longer than the age of the universe (Rasio
et al. 2001).

A second important tendency in binary-single interactions is that the final
binary is likely to consist of the two most massive of the three original
stars (e.g., Sigurdsson \& Phinney 1993).  Therefore, objects such as an
IMBH are probably commonly found in binaries, even if they were originally
single.  This has important implications for gravitational radiation, as we
discuss later.  The specific results of hardening depend on whether the
most massive stars are on the main sequence (for young clusters) or are
compact stellar remnants (for old clusters).  We now discuss each of these.

\subsection{Young stellar clusters}

If a stellar cluster is less than a few tens of millions of years old,
then the most massive stars are still on the main sequence.  Their sizes
are therefore significant, and collisions or mergers are possible.
Ebisuzaki et al. (2001) and Portegies Zwart \& McMillan (2002) have
proposed that in a young cluster, core collapse may lead to multiple
mergers of a single star with other stars, producing an extremely
massive star that will leave behind a black hole of several hundred solar
masses when the nuclear fuel of the star runs out.

The simulations backing this conclusion are not yet able to include the
effects of binaries, because the binary orbital timescale is so much less
than the relaxation time of  the cluster.  There are two expected effects
from binaries that go in  opposite directions.  The first is the injection
of heat mentioned above. This means that if there are many primordial
binaries (as expected for massive stars based on observations of current
star-forming regions; e.g., Elson et al. 1998), the central density of the
cluster will not be as large as it would for single stars.  By itself, this
would decrease the collision rate significantly, and would therefore
inhibit the growth of a large star by mergers.  However, binary-single
interactions are very complicated, meaning among other things that in such
an encounter, two stars can pass very close to each other. For example, for
three equal-mass stars the probability of some pair of stars passing within
a distance $\epsilon a$ or less of each other (where $a$ is the original
binary semimajor axis and $\epsilon<1$) is $\sim \epsilon^{1/2}$ (Hut 1984;
McMillan 1986; Sigurdsson \& Phinney 1993).  This tends to increase the
collision rate.

It is not yet clear which of these effects is more important.  It seems
likely that the overall rate of collisions is increased by the presence of
binaries.  However, for growth of a large star it is essential that the
large star undergo many collisions.  The probability of multiple collisions
for a single star may be decreased significantly by the decrease in stellar
number density produced by binaries, but this has to be  investigated.  In
addition, although the mergers themselves are likely to occur with little
mass loss (see Lai, Rasio, \& Shapiro 1993; Rasio \& Shapiro 1994, 1995),
the question is whether the mergers can happen rapidly enough to offset
mass loss by winds, pulsational instabilities, or stellar evolution.

\subsection{Old stellar clusters}

If a cluster is more than $\sim 10^8$~yr old, then the remaining main sequence
stars have masses $\lta 10\,M_\odot$ and are therefore less massive than
stellar-mass black holes.  If a cluster is more than a few billion years
old, then the main sequence stars are less massive than $\sim 1.5\,M_\odot$
neutron stars.  Therefore, in old clusters the most massive objects are
compact stellar remnants.  These compact objects have negligible cross
sections, hence direct collisions do not happen.  However, if the binaries
get tight enough, gravitational radiation may play a major role.

Kulkarni, Hut, \& McMillan (1993) and Sigurdsson \&  Hernquist (1993)
have examined whether gradual tightening of a black hole binary in a
cluster could lead to mergers in the cluster due to gravitational
radiation. The issue is that the change in binding energy of a binary in
an interaction  is proportional to the original binding energy (Heggie
1975), therefore the recoil kicks become stronger as the binary
tightens.  Kulkarni, Hut, \& McMillan (1993) and Sigurdsson \&  Hernquist
(1993) found that for interactions of a single $10\,M_\odot$ black hole
with a binary $10\,M_\odot-10\,M_\odot$ black hole, the recoil speed
exceeds the $\sim 50$~km~s$^{-1}$ escape velocity from the core of a
globular (Webbink 1985) before the binary becomes tight enough to merge
by gravitational interaction. This may mean that globulars are the
engines for many mergers, but that the mergers themselves happen well
outside globulars (Portegies Zwart \& McMillan 2002).  On the other hand,
Miller \& Hamilton (2002a) showed that if one of the black holes in the
binary has an  initial mass $M\gta 50\,M_\odot$, its inertia keeps it in
the cluster and therefore it can undergo repeated mergers and, in
principle, grow to $M\sim 10^3\,M_\odot$ or more.  In addition,
binary-binary interactions might produce hierarchical triples in which
the inner binary undergoes large eccentricity oscillations via the Kozai
mechanism, such that the inner binary can merge without any dynamical
recoil (Miller \& Hamilton 2002b).  The rates of such interactions, and
the efficiency with which a black hole can grow, need to be investigated
more thoroughly to determine the viability of this mechanism.

\section{Implications for Gravitational Radiation}

The rates and properties of gravitational waves from intermediate-mass
black holes depend on many unknowns, including the number density of IMBH
in the universe, the types of interactions they undergo, and their formation
process.  Their number density may be addressed by future X-ray and optical
data (see the contributions by Mushotzky and van der Marel in these
proceedings).  For example, observations in the next few years may clarify 
the fraction of globular clusters that harbor IMBH, and
the mass distribution of those black holes.

The formation process is also important.  In the Population III and young
cluster scenarios, the initial mass of the black hole is several hundred
solar masses.  These high masses mean that the frequency of gravitational
waves from IMBH in binaries would be at most a few Hertz, which is too low
for detectability from the ground.  However, the longer-term inspiral of,
e.g., a stellar-mass black hole into an IMBH in a cluster could be detected
with space-based instruments such as LISA.  If IMBH are common in globulars,
then there may be tens of globulars around the Milky Way with IMBH binaries
that have frequencies in the $10^{-4}~{\rm Hz}\lta f_{\rm GW}\lta 1~{\rm Hz}$
range of LISA (Miller 2002).  These are persistent sources, and therefore
the signal to noise will be increased by continued observation. However, the
majority of those will have frequencies $f_{\rm GW}<10^{-3}$~Hz, hence they
will suffer confusion with the unresolved white dwarf binary background. A
more promising source is globulars in the Virgo cluster of galaxies.
Although Virgo is $\sim 10^3$ times more distant than Galactic globulars, it
has several hundred times more globulars than our Galaxy, with the result
that a few IMBH binaries are likely to be detectable in a few years with
LISA (Miller 2002). The frequencies of detectable IMBH binaries in VIRGO
will be $f_{\rm GW}\gta 10^{-3}$~Hz, and therefore relatively free of
contamination.

If instead IMBH form in old clusters, their initial masses are likely to
be tens of solar masses (Miller \& Hamilton 2002a).  These masses imply
gravitational wave frequencies in merger and ringdown of a few tens of
Hertz, which is within reach of ground-based instruments.  As many
as tens of events per year could be detected with Advanced LIGO (Miller
2002).  The LISA rate would also be enhanced slightly,
because this scenario involves several mergers before becoming
indistinguishable from pictures in which the initial mass is high.

In either case, although by the time the binaries enter the bandpasses of
ground-based instruments they will be nearly circular (Wen 2002), in the LISA
band the binaries are likely to have large eccentricities.  This is because
three-body interactions will continue until the gravitational radiation
timescale is less than the time to the next encounter.  The strong dependence
of merger timescale on eccentricity (Peters 1964),
\begin{equation}
\tau_{\rm GW}\approx 3\times 10^8(M_\odot^3/\mu M^2)(a/R_\odot)^4
(1-e^2)^{7/2}~{\rm yr}
\end{equation}
for a total binary mass $M$ and reduced mass $\mu$, means that if a binary
is pushed to high eccentricity it is more likely to merge. This produces a
significant selection effect towards high eccentricities, and means that
the initial eccentricity of a binary on its final inspiral is usually
$e\gta 0.9$ (Gultekin, Miller, \& Hamilton, these proceedings).
Therefore, effects such as pericenter precession will be detectable in the
LISA signals.  For IMBH binaries in the VIRGO cluster, one expects to see
orbital decay as well (in addition, possibly, to Lense-Thirring
precession), which gives enough information to solve for the distance to
Virgo using just gravitational radiation (Miller 2002).

A final intriguing possibility for IMBH gravitational waves is that some
number of IMBH may merge with the central supermassive black hole of a given
galaxy (Madau \& Rees 2001).  Such events, which are the merger of a $\sim
10^3\,M_\odot$ black hole with a $\sim 10^{6-9}\,M_\odot$ black hole, have
extremely high mass ratios and therefore can be treated using test particle
techniques (Hughes 2001).  In addition, since IMBH have masses tens to
hundreds of times those of stellar-mass black holes, the signal to noise of
such a merger is much greater than that of a stellar-mass black hole with
a supermassive black hole.  An event such as this would therefore provide
incredibly precise probes of the spacetime of a rotating supermassive
black hole.  The event rate is difficult to estimate, but if tens of IMBH
fall into a typical supermassive black hole in a Hubble time then a few
to tens of such events per year can be expected (Miller 2002).

\section{Impact of Future Observations}

Despite the many exciting implications of intermediate-mass black holes,
we must keep in mind that at this point they are not conclusively 
established to exist.  Unambiguous measurements of the mass are required.
For the ultraluminous X-ray sources the only way to do this is to identify
a stellar companion and do radial velocity measurements, a very difficult
task for objects that distant.  For IMBH candidates
in globular clusters, however, there are hints of effects in current data
that may lead to dramatic improvements in our understanding.

Observations of M15 (Gerssen et al. 2002) and possibly other globulars (K.
Gebhardt, personal communication) suggest that the cores of these
globulars are rotating rapidly.  The evidence comes from radial velocity
measurements of stars and fits of models to them, and may soon be
supplemented by proper motion measurements (K. Gebhardt, personal
communication).  Taken at face value, the evidence suggests that in the
cores the ratio of rotation speed to velocity dispersion is $v_{\rm
rot}/\sigma\sim 1$.  This is a surprising result.  Simulations using
n-body codes suggest that this rotation will be communicated outward, and
that in a core relaxation time there will be little net rotation unless
there is a supply of angular momentum in the core (e.g., Einsel \&
Spurzem 1999).  Given that the
core relaxation time is as short as  $\sim 10^7$~yr in dense clusters such
as M15, how can this angular momentum be supplied?

One possibility invokes an IMBH in a binary with a stellar-mass black hole
(F. Rasio, personal communication).  Such a binary has the required angular
momentum, and can have the necessary rate (see Miller \& Colbert 2003 for a
more complete discussion).  In addition, because after the binary hardens
and merges the next binary would have an orbital plane at a random angle to
the previous one, the position angle of rotation in the cluster is expected
to wander as one measures farther from the center, consistent with
observations (Gerssen et al. 2003). If the observational interpretation is
confirmed, and if no other theoretical explanation is found, this has extremely
exciting implications for gravitational wave generation.  Not only would it
confirm that IMBH exist, it would show that right now, IMBH in globulars are
undergoing frequent coalescences with stellar-mass black holes, hence are
strong sources of gravitational radiation with unique potential as 
astrophysical probes and testing grounds for predictions of
general relativity.

{\bf Acknowledgements}

We are grateful to Kayhan G\"ultekin, Doug Hamilton, and Steinn
Sigurdsson for many enlightening conversations.  This work was supported in
part by NSF grant AST~0098436 and NASA grant NAG 5-13229.

\end{document}